\documentclass[prd,preprint,preprintnumbers,superscriptaddress,nofootinbib,11pt]{revtex4-1}
\usepackage[a4paper]{geometry}

\usepackage{amstext,amssymb}
\usepackage{amsmath}
\usepackage{graphicx}
\usepackage[hyperfootnotes=true]{hyperref}
\usepackage{xspace}
\usepackage{color}
\usepackage{slashed}
\usepackage{chngcntr}

\begin{document}

\title{Gravitational production of massive scalars in the context of inflation}
\author{Urjit A. Yajnik}
\email{yajnik@iitb.ac.in}
\affiliation{Tata Institute of Fundamental Research, Homi Bhabha Road, Bombay 400005, India}
\preprint{TIFR-TAP-11}
\preprint{July 1988}

\begin{abstract}
We set up a formalism for calculating the energy density generated in a quantized massive scalar field in the course of the drastic change in spacetime geometry at the end of the inflationary era. The calculation relies on the notion of adiabatic vacuum.
The Bogolubov coefficients are computed by employing the sudden approximation. After obtaining a general formula, we calculate explicitly the energy density generated in a particle species with $m/H \ll1$, where $m$ is the particle mass and $H$ is the Hubble constant during the inflationary epoch. We find the contribution of the long-wavelength modes to be $\propto H^5/m$. If such particles are very weakly interacting, they can come to dominate the total energy density in the Universe. Other cosmological implications are also discussed.\\

\vspace{2cm}\noindent
\begin{quote}
\textit{This preprint contains the calculations supporting the results published in Phys. Lett. B 234 (1990) 271-275. The unpublished preprint is now typeset in \LaTeXe and submitted to the arXiv due to the renewed attention it has received.
}
\end{quote}

\end{abstract}

\maketitle

\section{Introduction}\label{sec:intro}
Quantum field theory in curved space\cite{1}\cite{2}  is one of the most extensively explored approaches to a full quantum theory of gravity. And yet it has some very important issues
that remain unresolved. One is that of defining the vacuum state. Another is of computing the back reaction on the space in which quantization is carried out. It is therefore important to test whatever conceptual advances have been made so far by applying them to as many new situations as possible. Inflationary universe\cite{3} is a very interesting proposal whose correctness has not yet been established. While a search is on for particle physics models within which it may be realized, or for variants of the scenario in order to avoid pitfalls, it is also worth pursuing physical effects other than those the scenario was originally invented to produce (or to pre-empt). It is in this spirit that we carry out the present investigation.

The most dramatic effect of quantizing a field in a curved space is that in the generic case, quanta of the field must be spontaneously produced in the vacuum. This effect of course can occur for a field coupled to any classical background field. In the case of gravity, the situation is further complicated because the very concept of a particle, or equivalently, that of the vacuum state of the quantum field theory becomes ill-defined. Instead of worrying about particle number, one might construe the problem as that of exchange of energy between the quantized field and gravity. One would then try to compute the expectation value of the energy-momentum stress tensor for the field. But this attempt suffers from the dual handicap of the nonuniqueness of the ground state in which to compute such an expectation value, and the divergences typical of quantum field theory. So far, many techniques have been developed to deal with these problems. In this paper we use the adiabatic procedurea\cite{4} for defining the vacuum and a simple normal ordering prescription to calculate finite $\langle T_0^0 \rangle$. The latter prescription, to be discussed in \ref{sec:sec2}, is satisfactory for Friedmann- Robertson-Walker (FRW) universes but may not be satisfactorily generalizable to other cases.

Our interest in carrying out this calculation is guided by the fact that the production of particles is larger, the greater the rate of change of geometry is\cite{6}. Since the generic inflationary scenario involves a change, over a relatively short time scale, from a de Sitter like or inflationary phase to a radiation dominated Friedmann phase, we expect it to be accompanied by copious production of particles. The case of graviton production has been studied extensively in the past\cite{7}\cite{8}. On the other hand, production of massive particles has not received adequate attention.

While the present calculation was in progress, Turner and Widow\cite{9} have calculated the same effect, using a different approach. Their particles originate as quantum fluctuations during the inflationary phase. The evolution of these fluctuations is then traced as the wavelengths of the modes cross various critical physical scales. We shall compare their results with ours in the Conclusion.

Our formalism suggests two cases of interest: $m \ll H$, and $m \sim H$. We are able to display explicitly the dependence of the magnitude of the energy density on the mass only for the small mass case. In the case of $m \sim H$, we obtain the spectrum and the energy density for a representative value of $m$. Our main results are that particles of small mass contribute a large amount to the total energy density, which is comparable to the background energy density if, for instance, inflation is driven by $\rho_{0}\simeq M_{G U T}^{4}\sim(10^{14}G e V)^{4}$ and if $m\lesssim0.01$eV. For particles of larger mass, and which interact rapidly with other particles, we get a nontrivial contribution to the total energy density, but we may get no distinctly observable signature today. On the other hand, if these particles are very weakly interacting, they can come to dominate the total energy density at a later stage and cause a conflict with the nucleosynthesis data.

In deriving the above results however, we have not taken into account the effects of nonzero temperature. This may be important for the particles of standard model which acquire mass through spontaneous symmetry breaking, and which remain effectively
massless at high temperature. Thus it is unclear whether the results obtained here are of direct validity for the known particles, but the results are suggestive enough to warrant an independent study for the case of spontaneous symmetry breaking. Some more remarks are included in sec. \ref{sec:sec5}.

In what follows, in sec. \ref{sec:sec2} we describe the formalism to be used and set up the problem. In sec. \ref{sec:sec3} we present the calculation and the general result. In section \ref{sec:sec4} we use the formula to obtain the energy density spectrum $d\rho/d k$ and the total energy density of the produced particles. In section \ref{sec:sec5}, we consider the cosmological consequences of the results. Section \ref{sec:sec6} contains concluding remarks.
 
 \section{The Formalism}\label{sec:sec2}

Consider the field theory of a real massive scalar field in a spatially flat FRW universe. We shall write the line element for the spacetime in the conformal form
\begin{equation}
d s^2=\Omega^2(\eta)\left(-d \eta^2+|d \vec{x}|^2\right)
\label{eq:metric}
\end{equation}
Given this coordinatization, the theory of a massive scalar field is given by the action
\begin{equation}
A=\frac{1}{2} \int d \eta d^3 x \Omega^4\left\{\frac{1}{\Omega^2}\left(\frac{\partial \phi}{\partial \eta}\right)^2-\frac{1}{\Omega^2}|\vec{\nabla} \phi|^2-m^2 \phi^2\right\}
\label{eq:fieldaction}
\end{equation}
We have chosen minimal coupling to scalar curvature. There is no experimental evidence to the contrary, nor is there any theoretical motivation in the form of higher symmetry, (such as exists, for instance, in the massless case), to justify any nontrivial coupling.

In the quantum theory, we expand the field operator $\phi$ in terms of mode functions in the following form
\begin{equation}
\phi(\vec{x}, \eta)=\int \frac{d^3 k}{(2 \pi)^{3 / 2}}\left\{a_{\vec{k}} \frac{\chi_k(\eta)}{\Omega(\eta)} e^{-i \vec{k} \cdot \vec{x}}+a_{\vec{k}}{ }^{\dagger} \frac{\chi_k^*(\eta)}{\Omega(\eta)} e^{i \vec{k} \cdot \vec{x}}\right\}
\label{eq:fieldexpansion}
\end{equation}

With the given scaling, the $\chi_k$ satisfy (with prime denoting differentiation with respect to $\eta$)
\begin{equation}
\chi_k^{\prime \prime}-\frac{\Omega^{\prime \prime}}{\Omega} \chi_k+|\vec{k}|^2 \chi_k+m^2 \Omega^2 \chi_k=0
\label{eq:modeeqn}
\end{equation}
The system is canonically quantized with
\begin{equation}
\left[a_{\vec{k}}, a_{\vec{k}^{\prime}}^{\dagger}\right]=\delta^3\left(\vec{k}, \vec{k}^{\prime}\right)
\end{equation}
provided the Wronskian of the $\chi$-functions is normalized to
\begin{equation}
\chi_k \chi_k^{* \prime}-\chi_k^* \chi_k^{\prime}=i
\label{eq:Wronskian}
\end{equation}
For this system we shall compute
\begin{equation}
\left\langle\psi\left|T^0{ }_0\right| \psi\right\rangle=\left\langle\psi\left|\frac{1}{2 \Omega^2}\left\{\left|\phi^{\prime}\right|^2+|\vec{\nabla} \phi|^2+\Omega^2 m^2|\phi|^2\right\}\right| \psi\right\rangle
\label{eq:Tvev}
\end{equation}
The modulus sign refers to extracting real numbers from complex numbers and vectors in the standard way, and also the hermitian combination according to $|A|^2=A^{\dagger} A$ for operators. Here $|\psi\rangle$ is meant to be any appropriately chosen ground state.

We choose $|\psi\rangle$ by demanding
\begin{equation}
a_{\vec{k}}\ |\psi\rangle\ =\ 0
\label{eq:vacchoice}
\end{equation}
This choice of course is not unique. It was specified by assuming $\chi_{k}(\eta)/\Omega(\eta)$ to be ``positive frequency" mode functions. Due to the general covariance of the classical theory, another choice $\tilde{\chi}_{k}(\eta)=\,\alpha_{k}\chi_{k}(\eta)+\beta_{k}\chi_{k}^{*}(\eta)$, (appropriately normalized), would have been just as good. The corresponding $\tilde{a}_{\vec{k}}$ would yield a different vacuum than in Eq. (\ref{eq:vacchoice}). Although a FRW space offers a natural choice for $3+1$ slicing into hypersurfaces of constant $Tr K$ (trace of the extrinsic curvature tensor),\cite{10} this favoured choice of slicing is not available in arbitrary spacetime manifolds. In such cases, each possibility for slicing introduces further possibilities for the choice of a vacuum.

In the FRW case at hand, we shall make use of the natural choice it affords for slicing (and which is implied in the expression Eq. \eqref{eq:metric} for the metric). We are then left with having to decide on one particular linear combination from the set of two fundamental solutions of Eq. \eqref{eq:modeeqn}, which should be called the positive frequency mode function. In the language of complex vector spaces, this amounts to singling out a particular complex structure on the one-complex-dimensional space spanned by the two real solutions of \eqref{eq:modeeqn}. For this purpose we shall use the notion of adiabatic vacuum developed by Parker\cite{11} (see also ref. \cite{2} for other references). We shall not discuss the details of this approach here, though the prescription will be stated briefly in sec. \ref{sec:sec3} where it is used. Another natural way to define the vacuum in the conformally flat case such as we have, is the conformal vacuum prescription\cite{2}. This will not be used in the present work.
 
 To continue, we note that in any vacuum $|\psi\rangle$ suitably identified, Eq. \eqref{eq:Tvev} becomes
 \begin{equation}
 \langle\psi|T^0{ }_0|\psi\rangle=\int\!\frac{d^{3}k}{(2\pi)^{3}}\,\frac{1}{2\Omega^{2}}\,\left\{\left|\left(\frac{\chi_{\vec{k}}}{\Omega}\right)'\right|^{2}+\left(\frac{|\vec{k}|^{2}}{\Omega^{2}}+m^{2}\right)|\chi_{\vec{k}}|^{2}\right\} \left\langle \psi|a_{\vec{k}}^{\dagger}a_{\vec{k}}+a_{\vec{k}}a_{\vec{k}}^{\dagger}|\psi\right\rangle
 \label{eq:Tvevmodes}
 \end{equation}
 We have already dropped the off-diagonal terms that do not contribute to the vacuum expectation value. The remaining two terms will be together interpreted by normal ordering just like in flat space. That is, the operator expression for $T^0_0$ will be taken with $a_{\vec{k}}$'s to the right of $a^\dag_{\vec{k}}$'s, and the $c$-number generated in this arrangement will be discarded. Since this makes $\langle\psi|T_{\phantom{\mu}\nu}^{\scriptscriptstyle\mu}|\psi\rangle$ vanish for all
 $\mu$, $\nu$, the $c$-numbers constitute a tensor. Thus the normal ordering prescription is covariant, with the proviso that we do not seek a new
 complex structure for the quantum theory when we carry out a reparameterization of the
background manifold. For example, in the FRW case, a natural slicing is already suggested, so that the only reparameterizations affecting 
$\chi_k(\eta)$ are $\eta\rightarrow{\bar{\eta}}(\eta)$, which do not warrant
making a new choice for mode functions in which $\chi_k$ and $\chi^*_k$ get mixed. The only change needed (since the $\chi_k$ are scalar), is $\chi_{k}(\eta)\,\rightarrow\,\chi_{k}(\eta(\bar{\eta}))$, so that $\langle\psi|T_{\phantom{\mu}0}^{0}|\psi\rangle$
will continue to be zero. Finally, the natural $\chi_k$'s will be singled out for us by the adiabatic procedure. The answer produced by the latter procedure should not get affected by  $\eta\rightarrow{\bar{\eta}}(\eta)$ transformations.

Thus, in the symmetric case of FRW, we utilise the freedom allowed by the reparameterization gauge symmetry (viz., choose the preferred slicing) to obtain a sensible answer. However, in a manifold lacking symmetry, no preferred time direction will exist, and different observers choosing different coordinate systems for their convenience will also decide to choose different complex structures consistent with their time direction, and the proviso on reparameterization stated in the preceding paragraph will appear unjustifiable.

For the symmetric case of FRW, the above picture accords with our intuition. The free (although coupled to a classical field) field theory has a stable ground state of zero energy; a single particle state has energy
\begin{equation}
w_{k}(\eta)\quad\equiv\quad\Omega\mathrm{~}\left\{\left|\left(\frac{\chi_{k}}{\Omega}\right)^{\prime}\right|^{2}+\left(\frac{|\vec{k}|^{2}}{\Omega^{2}}+m^{2}\right)|\chi_{k}|^{2}\right\}
\label{eq:modefrequency}
\end{equation}
which changes with time $\eta$ due to interaction with the gravitational field, conventionally known as gravitational red or blue shifting.

 If a particular $|\psi\rangle$ is adhered to in a FRW universe as in the preceding discussion, no ``particle creation" ( or, in less picturesque language, spontaneous generation of matter energy density, gravitational in origin), can occur. However, the energy density as defined
in Eq. \eqref{eq:Tvevmodes} can change spontaneously if physical considerations indicate choosing different $|\psi\rangle$'s at different epochs. The simplest of such cases are the toy models in which the conformal scale factor $\Omega$ evolves from one constant value to another in a finite duration of
time\cite{12}\cite{6}. One is led to define two distinct vacua, one (say the ``in") during the beginning epoch of constant $\Omega$ and another, (say the ``out") during the final epoch of constant $\Omega$. Then the out vacuum turns out to be a state of infinite particle number built on the in
vacuum, and we would refer to such a phenomenon as creation of particles. The amount of particles produced, and their spectrum are determined to a large extent by the rate at which the scale factor changes in the transition region.

We expect inflation to lead to production of particles for similar reasons. The universe is supposed to have undergone a change from a de Sitter like phase to radiation dominated phase over a relatively short period of time. Any reasonable vacuum defined in the de Sitter like phase can turn out to be a state of nonzero particle number built on a vacuum
defined in the radiation dominated phase. The main difference from the toy models is that we do not have flat space in the past or in the future of the transition region. For this purpose we rely on the notion of adiabatic vacuum. We define adiabatic vacuum $|\psi^I\rangle$ in the de Sitter like or the inflationary phase, and $|\psi^R\rangle$  in the radiation dominated phase. We then need the Bogolubov transformations\cite{13} relating the two vacua, i.e., a set of transformations that relate the creation and destruction operators of the two phases
\begin{equation}
a_{\vec{k}}^{R}=\alpha_{k}a_{\vec{k}}^{I}+\beta_{k}a_{\vec{k}}^{I}
\label{eq:bogtran}
\end{equation}
In practice, this requires knowing the mode functions valid throughout the history of the universe, including the region of transition. Calculation of the Bogolubov coefficients can then be carried out as in the toy models mentioned above. Unfortunately, this is too
difficult in the present case, as will be discussed in the concluding section. Here we shall
adopt what may be called the \textit{sudden approximation}\cite{14}. We assume that the Universe goes
over from the inflationary phase to the radiation dominated phase abruptly. We then match the mode functions at the juncture of the two phases:
\begin{equation}
  \frac{\chi_{k}^{I}}{\Omega}=\alpha_{k}\frac{\chi_{k}^{R}}{\Omega}+\beta_{k}\frac{\chi_{k}^{R*}}{\Omega}\qquad   \mathrm{(at\  the\  juncture)}
  \label{eq:junctionfield}
\end{equation}
and, because the mode functions satisfy a second order differential equation,
\begin{equation}
  \left(\frac{\chi_{k}^{I}}\Omega\right)^{\prime}=\alpha_{k}\left(\frac{\chi_{k}^{R}}\Omega\right)^{\prime}+\beta_{k}\left(\frac{\chi_{k}^{R*}}\Omega\right)^{\prime}\qquad     \mathrm{(at\  the\  juncture)}
  \label{eq:junctionfieldderivative}
\end{equation}
Due to the field expansion \eqref{eq:fieldexpansion}, these are the coefficients needed in \eqref{eq:bogtran}. Since the mode
functions continue to be normalised according to \eqref{eq:Wronskian}, the $\alpha$ and $\beta$ satisfy
\begin{equation}
|\alpha_{k}|^{2}-|\beta_{k}|^{2}=1
\end{equation}

The number $\textstyle\sum_{k}|\beta_{k}|^{2}$ can be interpreted as the number of particles created in the course of the phase change\cite{6}. In the present case, we evaluate 
\eqref{eq:Tvevmodes} with $|\psi\rangle=|\psi^{I}\rangle$ but with the normal ordering of the radiation dominated phase. We then get
\begin{equation}
\langle\psi^{I}|T_{\;\;0}^{0}\,|\psi^{I}\rangle=\frac{1}{\Omega^{3}}\int\!\frac{d^{3}k}{(2\pi)^{3}}w_{k}|\beta_{k}|^{2}
\label{eq:Tzerozerovev}
\end{equation}
This is the contribution to the energy density in the radiation dominated phase, from particles gravitationally generated due to ending of inflation.

The procedure of mode matching for obtaining the Bogolubov coefficients has been used previously\cite{15}\cite{9} in the context of graviton production. The limitation of the approximation
is that we lose the information about the modes with wavelengths comparable to or smaller than the (temporal) length of the transition region. However, due to technical difficulties to be explained in the conclusion, this is the best we have been able to do. It should be noted however, that in any reasonable formalism (such as the exact adiabatic treatment of the entire evolution), the value of $|\beta_{k}|^{2}$ becomes exponentially damped for wavelengths smaller than the typical time scale involved in the transition. As such, the wavelengths we
have foregone the information about are not of much interest either.

\section{Adiabatic vacua and mode matching}\label{sec:sec3}
Let $\eta_1$ and $\eta_2$ denote the times when the inflationary phase begins and ends, respectively. In most inflationary universe models, the epoch between the Planck time, (before which we do no know the physics), and the time $\eta_1$, the Universe is radiation dominated. In the models with primordial inflation, the time n in fact runs into the Planck time. In either case, since inflation must enhance the scale factor by the stupendous factor of $10^{27}$ or more (see Guth\cite{3}), any relics of the time h and earlier must become completely irrelevant to physics. In particular, this allows us to assume that at the time $\eta_2$, our scalar
field of Eq. 
\eqref{eq:fieldaction} is in its ground state\footnote{The validity of this assumption is not at all obvious, especially because the dilution that occurs is of energy density but not, for instance, of the number of occupied modes per causal horizon volume. We adopt this assumption here primarily because it simplifies the calculation without sacrificing essential physics. Also, it appears reasonable that at least in the range of wavelengths that we shall be concerned with at time $\eta_2$ there will be no relics, because at time $\eta_1$ these wavelengths were inordinately small compared to the natural physical scale of that epoch. I thank T. Padmanabhan for pointing this out.}. We shall not have much occasion to refer to $\eta_1$ in the following.

In addition, we shall not be concerned in the present work with the subsequent major changes in the course of the evolution of the Universe, such as transition to a matter dominated epoch.

We model inflation by taking the scale factor to be \cite{9}
\begin{equation}
\begin{aligned}
\Omega & =\frac{\Omega_2}{2-\eta / \eta_2} & & \eta_1<\eta<\eta_2 \\
& =\Omega_2 \frac{\eta}{\eta_2} & & \eta>\eta_2
\end{aligned}
\end{equation}

The constants have been chosen to ensure that the scale factor itself as well as its first derivative are continuous across $\eta_2$. The continuity of the first derivative is required by Einstein's equations which imply $\left(\Omega^{\prime} / \Omega^2\right)^2 \propto \rho$, which in turn may reasonably be assumed to be continuous. Assuming any greater degree of smoothness goes beyond the requirements of the approximation we are making. With the above choice of constants, the Hubble constant during the inflationary phase is given by
\begin{equation}
H^2 \equiv \frac{1}{\left(\Omega_2 \eta_2\right)^2}=8 \pi G \rho_0
\label{eq:Hubblesquared}
\end{equation}
where $\rho_0$ is the constant energy density that drives inflation. In the new inflation it has the order of magnitude $\left(10^{14} \mathrm{GeV}\right)^4$ to $\left(10^{16} \mathrm{GeV}\right)^4$ derived from Grand Unified Theories.

We now proceed to finding the mode functions and the adiabatic vacuua. In the de Sitter like phase, the mode functions satisfy
\begin{equation}
\chi_k^{\prime \prime}+\left\{-\frac{2}{\left(2 \eta_2-\eta\right)^2}+k^2+\frac{m^2 \Omega^2 \eta_2^2}{\left(2 \eta_2-\eta\right)^2}\right\} \chi_k=0
\end{equation}
The two independent solutions of this equation are found to be
\begin{equation}
\begin{aligned}
\chi_k & \sim\left(2 \eta_2-\eta\right)^{1 / 2} H_\lambda^{(1)}\left(k\left(2 \eta_2-\eta\right)\right) \\
& \sim\left(2 \eta_2-\eta\right)^{1 / 2} H_\lambda^{(2)}\left(k\left(2 \eta_2-\eta\right)\right)
\label{eq:Hankelmodes}
\end{aligned}
\end{equation}
with
\begin{equation}
\lambda^2=\frac{9}{4}-m^2 \Omega_2^2 \eta_2^2
\end{equation}
Here $H_\lambda^{(1)}, H_\lambda^{(2)}$ are Hankel functions of order $\lambda$. The adiabatic prescription requires us to identify that combination (and normalisation) of the fundamental set \eqref{eq:Hankelmodes} to be the positive frequency mode which behaves as
\begin{equation}
\chi_k \sim \frac{1}{\sqrt{2 \bar{\omega}(\eta)}} \exp \left(-i \int^{\prime} \bar{\omega}\left(\eta^{\prime}\right) d \eta^{\prime}\right)
\end{equation}
in some limiting value of some parameter such as $m \Omega_2 \eta_2$ or $k$. The $\bar{\omega}$ appearing in the ansatz is in turn to be obtained as the limit of
\begin{equation}
\omega^2 \equiv-\frac{2}{\left(2 \eta_2-\eta\right)^2}+k^2+\frac{m^2 \Omega^2 \eta_2^2}{\left(2 \eta_2-\eta\right)^2}
\end{equation}
in the same limit of the same parameter. For the case at hand, the answer appears in \cite{2}, sec.5.4, which one can check\footnote{Note that in \cite{2}, the argument of the Hankel function $k \eta=-(k / H) e^{-H t}$ increases as the Universe expands, whereas our argument $k\left(2 \eta_2-\eta\right)$ decreases. Hence our positive frequency function is $H^{(1)}$.}.
\begin{equation}
\chi_k^{I(+)}(\eta)=\frac{1}{2}\left(\pi\left(2 \eta_2-\eta\right)\right)^{1 / 2} H_\lambda^{(1)}\left(k\left(2 \eta_2-\eta\right)\right)
\label{eq:infpositivemode}
\end{equation}
and similarly the negative frequency function in terms of $H^{(2)}$.


In the radiation dominated phase one finds
\begin{equation}
\chi_k^{\prime \prime}+\left(k^2+\frac{m^2 \Omega_2^2}{\eta_2^2} \eta^2\right) \chi_k=0
\label{eq:radmodeeqn}
\end{equation}
The fundamental solutions for this are the parabolic cylinder functions of Weber\cite{16}
\begin{equation}
\begin{aligned}
\chi_k & \sim D_\nu\left(e^{i \pi / 4} \sigma \eta\right) \\
& \sim D_\nu\left(e^{-i \pi / 4} \sigma \eta\right)
\label{eq:radmodes}
\end{aligned}
\end{equation}
with
\begin{equation}
\sigma=\sqrt{\frac{2 m \Omega_2}{\eta_2}} ; \quad \nu=-i \frac{k^2}{\sigma^2}-\frac{1}{2}
\end{equation}

We now note that the limit of the quantity in the brackets in \eqref{eq:radmodeeqn}, as either $\eta$ or $\sigma$ go to infinity is $\frac{1}{4} \sigma^4 \eta^2$, so that we must identify positive frequency modes to be that combination of \eqref{eq:radmodes} which behaves like $(\sigma \sqrt{\eta})^{-1} \exp \left(-i \frac{1}{4} \sigma^2 \eta^2\right)$ in the same limit. Using the fact that \cite{16}
\begin{equation}
D_n(z) \xrightarrow{|x| \rightarrow \infty} e^{-\frac{1}{4} z^2} z^n \quad \text { for }|\arg z|<\frac{3 \pi}{4}
\end{equation}
and using the Wronskian \cite{17}
\begin{equation}
\mathcal{W}\left\{D_n(z), D_{-n-1}(-i z)\right\}=e^{i \pi(1+n) / 2},
\end{equation}
we identify the radiation dominated era mode functions to be
\begin{equation}
\chi_k^{R(+)}(\eta)=\frac{e^{-\pi k^2 / 4 \sigma^2}}{\sqrt{\sigma}} D_\nu\left(e^{i \pi / 4} \sigma \eta\right)
\label{eq:radpositivemode}
\end{equation}
and similarly the negative frequency function in terms of the complex conjugate function.


We now match the mode functions at $\eta=\eta_2$. We get
\begin{equation}
\frac{1}{2} \sqrt{\pi \eta_2} H_\lambda^{(1)}\left(k \eta_2\right)=\frac{1}{\sqrt{\sigma}} \exp \left(-\pi k^2 / 4 \sigma^2\right)\left\{\alpha_k D_\nu\left(e^{i \pi / 4} \sigma \eta_2\right)+\beta_k(\text { c.c. })\right\}
\end{equation}
and
\begin{equation}
\begin{aligned}
\frac{\sqrt{\pi}}{2}\left\{\frac{-1}{\Omega_2 \eta_2}\right. & \left.\sqrt{\eta_2} H_\lambda^{(1)}\left(k \eta_2\right)+\frac{1}{2 \Omega_2 \sqrt{\eta_2}} H_\lambda^{(1)}\left(k \eta_2\right)+\frac{k \sqrt{\eta_2}}{\Omega_2} H_\lambda^{(1) \prime}\left(k \eta_2\right)\right\} \\
= & \frac{1}{\sqrt{\sigma}} \exp \left(-\pi k^2 / 4 \sigma^2\right)\left[\alpha _ { k } \left\{-\frac{1}{\Omega_2 \eta_2} D_\nu\left(e^{i\pi/4} \sigma \eta_2\right)\right.\right. \\
& \left.\left.+\frac{e^{i \pi / 4} \sigma}{\Omega_2} D_\nu^{\prime}\left(e^{i \pi / 4} \sigma \eta_2\right)\right\}+\beta_k\{\text { c.c. }\}\right]
\end{aligned}
\end{equation}
In both these equations (c.c.) means the complex conjugate of the coefficient of the preceding $\alpha_k$, and the prime on the special functions denotes derivative with respect to their own argument, not simply $\eta$. Now denote
\begin{equation}
\begin{aligned}
E & \equiv D_\nu\left(e^{i \pi / 4} \sigma \eta_2\right), \\
F & \equiv D_\nu^{\prime}\left(e^{i \pi / 4} \sigma \eta_2\right), \\
X & \equiv H_\lambda^{(1)}\left(k \eta_2\right) \\
Y & \equiv H_\lambda^{(1) \prime}\left(k \eta_2\right)
\end{aligned}
\end{equation}
and the complex conjugates by $\bar{E}, \bar{F}$ etc. After some simplification, and using the fact that due to the normalisation 
\eqref{eq:Wronskian},
\begin{equation}
\exp \left(-\pi k^2 / 2 \sigma^2\right)\left(e^{-i \pi / 4} E \bar{F}-e^{i \pi / 4} \bar{E} \boldsymbol{F}\right)=i
\end{equation}
we get a master formula
\begin{equation}
\begin{aligned}
\left|\beta_k\right|^2= & \frac{\pi}{4} \frac{\exp \left(-\pi k^2 / 2 \sigma^2\right)}{\sigma \eta_2}\left[\sigma^2 \eta_2^2 F \bar{F} X \bar{X}+E \bar{E}\left|\frac{1}{2} X+k \eta_2 Y\right|^2\right. \\
& \left.+2 \operatorname{Re} e^{i \pi / 4} \sigma \eta_2 F \bar{E} X\left(\frac{1}{2} \bar{X}+k \eta_2 \bar{Y}\right)\right]
\label{eq:masterformula}
\end{aligned}
\end{equation}


\section{Calculation of energy density}\label{sec:sec4}
\subsection{General discussion}\label{sec:sec4Gen}

Formula \eqref{eq:masterformula} gives the desired results in terms of standard special functions. It can be used to numerically plot $\left|\beta_k\right|^2$ as a function of $k$ given any value of $m$ (in the units of $H$ ). Instead, here we shall try to obtain analytic estimates of the dependence of $\left|\beta_k\right|^2$ on $k$ as well as on $m$. Hence we begin by identifying the different cases of interest.

One important scale in the problem is $H$; for two reasons. Firstly we recall that our approximation restricts us to values of $k_{phy} \equiv k / \Omega<H$. Secondly, on physical grounds we know that at $k_{p h y}>H$, the spectrum of produced particles gets exponentially cut off. The other scale to compare $k_{p h y}$ to is provided by $m$. We expect qualitatively different behaviours when $k_{p h y}$ is greater than or less than $m$ (i.e., when $\lambda_{p h y} \equiv 2 \pi / k_{p h y}$ is less than or greater than the Compton wavelength). This is at least true of flat space physics, where field theoretic phenomena set in when the system is probed at wavelengths small compared to the Compton wavelength. Considering both the above mentioned length scales together, we see that three cases arise. i) When $k_{p h y}$ crosses $m$ when it is yet smaller than $H$, ii) when it crosses $m$ when it is roughly $H$, and iii)when it does so beyond $H$. In other words, i) $m / H<1$, ii) $m / H \simeq 1$ and iii) $m / H>1$. Case ii) is distinguished from iii) only by the fact that the interesting range $k_{p h y} \simeq m$ is not lost in the exponential tail of the spectrum. Unfortunately, due to our approximation becoming unreliable precisely at $k_{phy} \simeq H$, the only information we can possibly get from \eqref{eq:masterformula}, both for cases ii) and iii) is restricted to the case $k_{p h y} \ll m$. But even this is difficult in the general case for technical reasons, i.e., it has to be done numerically. It turns out however, that we can get analytic estimate of the $k$ dependence even in this case for a particular value of $m$. This we shall do later. At first we take up the case $m\ll H$.

\subsection{The case of small mass}\label{sec:sec4SmallMass}

We take up this case first, partly because we will be able to do full justice to it by analytical methods, but also because it is the most interesting. One can learn the behaviour of $\left|\beta_k\right|$ for both the ranges $k_{phy} \ll m$, and $k_{phy} \gg m$ within the validity of our approximation, i.e., from \eqref{eq:masterformula}. Let us define spectral distribution of the energy density,
$d \rho / d k$ through
\begin{eqnarray}
\left\langle\psi^{R}\left|T^{0}_{\phantom{\mu}0}\right| \psi^{R}\right\rangle & =&\frac{1}{\Omega_{2}^{3}} \int \frac{d^{3} k}{(2 \pi)^{3}} w_{k}\left|\beta_{k}\right|^{2} \\
& \equiv& \int d k_{phy} \frac{d \rho}{d k_{p h y}}
\end{eqnarray}
The quantities $\left|\beta_{k}\right|^{2}$ and $w_{k}$ can be calculated as shown in Appendix \ref{sec:appA}. We find that immediately at the end of inflation,
\begin{eqnarray}
\frac{d \rho^{(1)}}{d k_{p h y}} & =&\frac{1}{8 \pi^{2}} H^{4}\left(\frac{1}{k_{p h y}}+\frac{H^{2}}{2 k_{p h y}^{3}}\right)\qquad  k_{p h y}^{2} \gg m H
\label{eq:lowmassspectrum-1}
\\
& =&0.04 \frac{H^{5}}{m} \frac{1}{k_{p h y}}  \qquad \qquad \qquad \qquad
k_{p h y}^{2} \ll m H
\label{eq:lowmassspectrum-2}
\end{eqnarray}
The superscript (1) is meant to label the case $m \ll H$. We find that the large $k$ case becomes independent of the mass. We then expect it to be the same as that, for instance, for gravitons \cite{9}. Indeed Appendix \ref{sec:appA} shows that our $\left|\beta_{k}\right|^{2}$ is the same as that obtained by Allen. However, our definition of the single-particle energy $w_{k}$, derived from our covariant definition of $\left\langle
T^0{ }_0\right\rangle$ differs from that used by Allen, which is simply $k_{phy}$.

The low-mode-number end of \eqref{eq:lowmassspectrum-2} shows a curious dependence on mass. As a result, for small mass, we can get a large contribution to the total energy density, coming from the extremely long wavelengths. We can estimate this contribution by integrating \eqref{eq:lowmassspectrum-2}. We need to put reasonable limits. We may take the upper limit to be $m$.

To determine the lower limit, recall the comment made at the beginning of sec. \ref{sec:sec3}, along with the footnote. At the epoch $\eta_{1}$ when inflation begins, the Hubble value is $H$ which sets the physical scale. Our entire calculation applies only to those modes that have $k_{p h y}\left(\eta_{1}\right)<H$. The reason is that the modes $k_{p h y}\left(\eta_{1}\right) \gtrsim H$ carry the information regarding the preceding (possibly radiation dominated) phase, and it is inappropriate to define the de Sitter adiabatic vacuum for them. Only for $k_{\text {phy }}\left(\eta_{1}\right) \ll H$, i.e., modes which do not explore the global geometry at $\eta_{1}$ can we safely pretend that the Universe has been truly de Sitter (with infinite past) and also that these modes are unoccupied at $\eta_{2}$. Hence at $\eta_{2}$, we take the lower limit on $k_{phy}$ as $\Omega_{1} H / \Omega_{2}$. We then get
\begin{equation}
\rho_{\text {low }}^{(1)}=0.04 \frac{H^{5}}{m} \ln \left(\frac{m}{H} \frac{\Omega_{2}}{\Omega_{1}}\right)
\label{eq:lowmassspectrumintegrated-1}
\end{equation}
Recall that inflation requires $\Omega_{2} / \Omega_{1} \sim 10^{27}$. This means that the low-mode-number
contribution vanishes if $m$ (or more generally, the upper limit of validity of Eq. \eqref{eq:lowmassspectrum-2}) becomes as small as $10^{-27} H$. Thus the lower limit on $k_{p h y}$ values dictated by the validity of the approximation scheme also sets a lower limit on the mass value for which the contribution Eq. \eqref{eq:lowmassspectrumintegrated-1} can be taken seriously.

We may similarly evaluate the contribution of high-mode-number modes. Here we take the upper limit to be $H$ and the lower limit, for the sake of argument, $\sqrt{m H}$. Then
\begin{equation}
\rho_{h i g h}^{(1)}=\frac{H^{4}}{8 \pi^{2}}\left\{\frac{1}{2} \ln \frac{H}{m}+\frac{H}{4 m}-\frac{1}{4}\right\}
\label{eq:massspectrumintegrated-2}
\end{equation}
The dominant contribution appears to be $\sim H^{5} / m$ as in the low-mode-number case, due to our rather liberal lower limit. If we ignore this in Eq. \eqref{eq:massspectrumintegrated-2}, we get a contribution essentially $\sim H^{4}$.

The case of massless scalars was considered by Ford \cite{18} and the related case of gravitons by Allen \cite{8}. It is interesting to check whether the zero-mass limit of our results agrees with their results. Eq. \eqref{eq:lowmassspectrum-1}, aside from its validity range indicated there for the small-mass case, is also the spectrum for the massless case. (Eq. \eqref{eq:lowmassspectrum-2} refers to a range of $k$ values not possible for the massless case). Our $\left|\beta_{k}\right|^{2}$ for this case 
\eqref{eq:A18} is the same as that obtained by above authors, and the first term of eq. \eqref{eq:lowmassspectrum-1} is the answer obtained by them. We additionally get the second term, due to our definition of $\rho$ which we have identified with $\left\langle T^{0}{ }_{0}\right\rangle$. Ford, Allen as well as Starobinsky compute $\rho$ by assigning to each mode of mode-number $k$ a single-particle energy $\hbar \omega_{k} \equiv \hbar k_{p h y}$. By contrast, our choice amounts to (see eq.s \eqref{eq:modefrequency}, \eqref{eq:A19} and \eqref{eq:A26})

\begin{equation}
\hbar w_{k}=\hbar\left(k_{p h y}+\frac{H^{2}}{2 k_{p h y}}\right)
\end{equation}
The definition of energy density in curved space is intrinsically ambiguous. We have here tried a choice that is justified by being a component of a covariant object. Further comments are included in the Conclusion.

\subsection{ Spectrum for one $m \sim H$ example }\label{sec:sec4msimH}
In this subsection we briefly mention the results for a particular value of $m$ close to $H$. We see that if $m^{2} \Omega_{2}^{2} \eta_{2}^{2}=\frac{1}{4} \sigma^{4} \eta_{2}^{2}=2$, the order $\lambda$ of the Hankel functions of (3.4) becomes $1 / 2$, in which case $H_{1 / 2}(z) \sim z^{-1 / 2} e^{i z}$ and the problem simplifies somewhat. The details are given in Appendix \ref{sec:appB}.
\begin{equation}
\frac{d \rho^{(2)}}{d k_{p h y}}=3.86 H^{2} k_{p h y} \quad m=\sqrt{2} H; \quad k_{p h y}^{2} \ll m H
\label{eq:fourpointsix}
\end{equation}
The superscript (2) denotes the case $m \sim H$. Note that the case $k^{2} / \sigma^{2} \gg 1$ is inappropriate here because with $m \sim H, \quad k_{p h y}^{2} \gg m H$ lends us far out of the domain of validity of our approximation. The so called low end $\left(k^{2} / \sigma^{2} \ll 1\right)$ may now be safely taken to be the representative of the entire allowed range $\left(k_{p k y}<H\right)$ of $k$.

We may take the above spectrum as representative of the case ii), i.e., $m \sim H$. A different value of mass will only change the numerical constant. We now integrate this spectrum, with upper limit $f H, f$ being some number less than one, since assuming this spectrum to be valid all the way to $\mathrm{H}$ is incorrect.
\begin{equation}
\rho^{(2)}=\mathrm{O}(1) \cdot f^{2} H^{4}
\label{eq:fourpointseven}
\end{equation}

\section{Implications for the Universe}\label{sec:sec5}
The expressions \eqref{eq:lowmassspectrum-1}, \eqref{eq:lowmassspectrum-2} and \eqref{eq:fourpointseven} give the energy density added to the background energy density immediately after inflation. Before considering the effects of this in the early Universe, we must make two caveats. The first concerns the fact that we have considered only a scalar field. whereas no fundamental massive scalar is known to date. This brings us to the second point. In the standard model. the known light particles acquire mass only after the temperature has dropped below the Weinberg-Salam symmetry breaking scale. It is unclear how this temperature dependent effect is to be taken account of in our formalism. One might suspect however, that this gravitational phenomenon occurring over length scales $H^{-1}$ may not be seriously affected by the temperature, which is $\sim M_{GUT} \gg H$.

The first thing we would like to check is whether the energy density of the produced particles is comparable to the background energy density. If it is, then we cannot really trust the answer because we have not taken account of the back reaction of this newly generated energy density on the background geometry. This problem is generic, and there is no known extension of the formalism to take this into account. Under the circumstances, in the case in which we find the gravitationally generated energy density to be comparable to the background density, we shall conclude that something nontrivial is occurring but which needs to be investigated using techniques yet to be invented.

In the following, while making order of magnitude estimates, we take the GUT mass scale $M_{G V T} \simeq \rho_{0}^{1 / 4}$ (see \eqref{eq:Hubblesquared}) to be $10^{14} \mathrm{GeV}, G^{-1 / 2} \equiv M_{P l} \simeq 10^{19} \mathrm{GeV}$, and $H^{2} \sim$ $\rho_{0} / M_{P l}^{2}=\left(10^{9} \mathrm{GeV}\right)^{2}$. First consider the terms in $\rho^{(1)}$ and $\rho^{(2)}$ that are $\sim H^{4}$. Substituting the above numbers, we see that these are certainly small compared to $\rho_{0}$. The interesting term is in $\rho_{\text {low }}^{(1)}$, which, aside from a logarithm is $\sim H^{5} / \mathrm{m}$. We find
\begin{equation}
\frac{\rho^{(1)}}{\rho_{0}} \sim \frac{H^{3}}{m M_{P l}^{2}}=
\frac{10^{-11} GeV}{m}
\end{equation}
This means that for $m \lesssim 0.01 \mathrm{eV}$, (and $m / H \lesssim 10^{-20}$) the formalism of quantum field theory in curved spacetime (in the form we have used here) breaks down. In fact if $\rho_{0}$ were as large as $\left(10^{16} \mathrm{GeV}\right)^{4}$ (so that $H \sim 10^{13} \mathrm{GeV}$ ), as in some models of inflation, this value would be $m \sim 100 \mathrm{eV}$.

Let us assume next that $m$ is larger than the limits found in the preceding paragraph, so that our formalism is still valid. First consider the case in which these particles are freely interacting with other species. Then they will quickly reach thermal equilibrium with the radiation constituting the background energy density. In that case their subsequent evolution is the same as that of a massive species in the standard cosmology. Despite the fact that they contribute significant fraction of the energy density at that epoch, there will be no observable signature left. It is possible that the produced particles decay into lighter particles. Then their evolution has to be traced as has been done in some dark matter scenarios \cite{19}.

The other possibility is that these particles are stable and interact very weakly with the rest of matter. Further, due to their low number density, their self-interaction may also be ignorable. In this case, the spectral distribution of this energy-density will remain
non-thermal. We then expect that the long-wavelength part $\left(k_{p h y}<m\right)$ can be treated as pressureless dust even as early as the epoch
$\eta_{2}$. Then these particles would come to dominate the total energy density too quickly, as follows
\begin{eqnarray}
\frac{\rho^{(1)}}{\rho_{\gamma}}(\eta) & \simeq &\frac{H^{5}}{m} \frac{T^{3}(\eta)}{T_{2}^{3}} \frac{1}{H^{2} M_{P l}^{2}} \frac{T_{2}^{4}}{T^{4}(\eta)} \nonumber \\
& \simeq &\frac{T_{2}}{m T(\eta)} \cdot 10^{-11} \mathrm{GeV}
\end{eqnarray}
If we take the reheating temperature $T_{2}$ to be $10^{14} \mathrm{GeV}$, for any $m \lesssim 10^7 \mathrm{GeV}$, the Universe would become matter dominated as early as the epoch of nucleosynthesis and affect that process adversely. (we took $T_{nucleosynthesis}=0.1 \mathrm{MeV}$ ). Thus the known abundance of primordial light nuclei places a lower limit on the mass of non-interacting particles that may be produced by this mechanism. As before, the bound is more stringent if $M_{GUT} \sim 10^{15}-$ $10^{16} \mathrm{GeV}$. In a more detailed treatment, we must remember that the actual contribution to the long-wavelength part of the spectrum is continuously increasing due to cosmological red-shifting.

It is clear that one needs to investigate the above possibilities in specific models of inflation and particle physics in greater detail. This will be reported in a separate publication.

\section{Conclusion }\label{sec:sec6}

We considered a quantised scalar field in an inflationary universe. Using some standard techniques, we calculated the energy density appearing in the form of quanta of this field due to the rather abrupt change from a de Sitter like phase to a radiation dominated phase. We employed the sudden approximation to do this calculation. This restricts the validity of our results to the case of modes with wavelengths long compared to the interval over which the transition from one phase to the other occurs. This is not as great a disadvantage as it might appear at first sight, since from other tractable problems we know that the spectrum of produced particles must taper off exponentially at large wave-numbers. The interesting case lost is that of $m \sim H$ for its modes with $k_{p h y} \sim m$. For small mass case we got reasonable results even for $k_{p h y} \sim H$. viz.. our $\left|\beta_{k}\right|^{2}$ is the same as that for the massless rase, calculated independently by Allen.

One may wonder whether a better treatment avoiding the sudden approximation was possible. I shall give here the problems encountered in the process. Ideally one would like to determine the mode functions that would be valid throughout spacetime, including through the transition region. One can then compare them with the adiabatic modes of the inflationary and the radiation dominated phases to obtain the required Bogolubov coefficients. To do this, we have to first model inflation using a smooth function that changes over from the $1 /(1-\eta)$ form to a linear function. With any function that does this, (that I could find), the differential equations satisfied by the mode functions no longer remain within the hyperbolic scheme. Although solutions may be still possible to find, one no longer has the advantage of the systematic approach possible with the hyperbolic scheme. Since the price paid in making the approximation is not great we have opted to rely on the sudden approximation.

The surprising result that comes out of this analysis is that the very long wavelength modes of the small mass particles give a large contribution to the energy density, proportional to $\mathrm{H}^{5} / \mathrm{m}$. We note that two factors contribute to this. One is 
\eqref{eq:A13}

\begin{equation}
\left|\beta_{k}\right|^{2} \simeq 0.74 \sqrt{\frac{H}{2 m}}\left(\frac{H}{k_{p h y}}\right)^{3} \quad k_{p h y}^{2} \ll m H
\end{equation}

The other is 
\eqref{eq:A23}

\begin{equation}
u_{k} \simeq 1.5 \sqrt{\frac{H}{2 m}} H \quad k_{p h y}^{2} \ll m H
\end{equation}

The appearance of $(H / m)^{1 / 2}$ in these expressions is puzzling. It has to do with how the modes of a massive scalar have to be normalised in a radiation dominated universe. Note the appearance of $\sigma^{-1 / 2} \sim m^{-1 / 4}$ in eq. 
\eqref{eq:radpositivemode}. This normlisation has appeared in the literature before. for instance, in ref. \cite{12}. As for the singular $k$ dependence of $\left|\beta_{k}\right|^{2}$. a more physical reasoning may exist, having to do with global properties of the two spaces concerned. Such a reasoning has been given by Allen in the massless case.

The singular $m$ dependence of the above equations need not worry us about the massless case because they are valid only for physical wavelengths larger than the Compton wavelength. and not appropriate for studying the massless limit. In this context. note that the mode functions we have found in eq.s 
\eqref{eq:infpositivemode}, 
\eqref{eq:radpositivemode} reduce to those for the massless case
as found, for instance, by Allen. The limit of the de Sitter era mode functions is obtained easily. For the mode functions of the radiation dominated era, one needs to consider $D_{n}(z)$ in the limit of $z\rightarrow 0$, and $n \rightarrow \infty$, but with $z^{2} n \rightarrow k^{2} \eta^{2}$. Such limits however, can not be taken after mode-matching has been performed at a specific epoch.

The massless limit can be found in our answer from the $k_{p h y} \sim H$ case. which turns out to be $O\left(m^{0}\right)$. In this case we find for $w_{k}$ (See eq.
\eqref{eq:A26})
\begin{equation}
w_{k}=k_{p h y}+\frac{H^{2}}{2 k_{p h y}} \quad k_{p h y} \lesssim H
\end{equation}
The second term here happens to be the same term from the complete expression for $w_{k}$ (eq. \eqref{eq:A19}
)that gave the leading contribution to the $k_{p \text { hy }}^{2} \ll m H$ case above. As pointed out below eq. 
\eqref{eq:massspectrumintegrated-2}, this term, although dictated by covariance, is not included in other works the author is aware of (viz., ref.s \cite{7}, \cite{8}, \cite{18}), in their definition of the effective energy of each mode. Energy-momentum tensor for gravitons is a subtle issue, but for massless non-conformal scalars the energy density must contain the above term, and to that extent we have corrected the result of ref. \cite{18}. Note that the above term becomes more important than the preceding one when $k_{p h y}<H$, but this is precisely the range in which modes are produced by the gravitational mechanism, and hence is not ignorable. The term is absent for conformally coupled fields. We plan to discuss this issue in greater depth in a separate publication dealing with the massless case.

We must also note that the present calculation has been carried out for scalar particles. The results however, are interesting enough that it seems worth carrying out a similar exercise for fermions. Otherwise, in absence of fundamental scalars, our results remain only of instructional value. Similarly, it is unclear what modifications, if any, the results will undergo if the finite temperature symmetry restoration effect of spontaneously broken gauge theories is taken into account.

The techniques of quantum field theory in curved space that we have used here are not well established. They are not unambiguous generalisations of the flat space field theory, and there has been no experimental conformation for them, for instance the notion of adiabatic vacuum. This calculation shows one example where the results may be testable against astrophysical observations. Since our procedure is not unambiguous, it is interesting to compare our results with those obtained by Turner and Widrow using a more
intuitive picture and heuristic arguments. Restated in the notation used here, their results for minimally coupled scalar massive particles are as follows. If we assume the values of $\rho_{0}$ and $T_{2}$ as above, their answer for the ratio of the energy density in these particles today to today's closure density is
\begin{equation}
\Omega_{\phi} \simeq 3.9 \times 10^{17} \cdot\left(\frac{H}{M_{P I}}\right)^{2} \cdot\left(\frac{m}{G e V}\right)^{\frac{1}{2}}
\end{equation}
in the range of $m$ values that coincides with that permitted by our formalism. If we take $m \simeq 10^{6} \mathrm{GeV}$ (which case we found conflicts with nucleosynthesis in our calculation if the particles are very weakly interacting), we get $\Omega_{\phi} \sim O(1)$, i.e.capable of seriously affecting the observed Universe as in our case. However, the dependence on $H$ and $m$ are completely different and we have no way of reconciling the two answers.

\begin{acknowledgments}
I thank Prof. J. V. Narlikar, Prof. S. M. Chitre, T. R. Seshadri, T. P. Singh and Ashoke Sen for useful conversations in the course of this work, and T. Padmanabhan for many useful discussions. I also wish to record here my indebtedness to Prof. P. Candelas
who taught me mathematical physics.

Note added in 2024 : This conversion of the original preprint has benefited from use of the tools at \url{mathpix.com}
\end{acknowledgments}


\appendix
\section{}\label{sec:appA}

The most important simplification that occurs on assuming $m / H\ll1$, that is, $m \Omega_2 \eta_2=\frac{1}{2} \sigma^2 \eta_2^2\ll1$ is that the order $\lambda$ of the Hankel functions of 
eq. \eqref{eq:infpositivemode} becomes $3 / 2$, in which case
\begin{eqnarray}
X=H_{3 / 2}^{(1)}\left(k \eta_2\right)=\sqrt{\frac{2}{\pi k \eta_2}}\left(-1-\frac{i}{k \eta_2}\right) e^{i k \eta_2} \\
Y=H_{3 / 2}^{(1) \prime}\left(k \eta_2\right)=\sqrt{\frac{2}{\pi k \eta_2}}\left\{\frac{3}{2 k \eta_2}-i+\frac{3 i}{2 k^2 \eta_2^2}\right\} e^{i k \eta_2}
\end{eqnarray}

Further, the argument $\sigma \eta_2$ of the parabolic cylinder functions goes to zero, in which case
\begin{equation}
D_\nu(0)=\sqrt{\pi} \frac{2^{\nu / 2}}{\Gamma\left(\frac{1-\nu}{2}\right)}
\end{equation}
\begin{equation}
D_\nu^{\prime}(0)=-\sqrt{\pi} \frac{2^{(\nu+1) / 2}}{\Gamma\left(-\frac{\nu}{2}\right)}
\end{equation}

It follows that
\begin{equation}
F \bar{F} X \bar{X} \sigma^2 \eta_2^2=2 \sqrt{2} \frac{\sigma^2}{k} \eta_2\left(1+\frac{1}{k^2 \eta_2^2}\right) \frac{1}{\left|\Gamma\left(\frac{1}{4}+\frac{i k^2}{2 \sigma^2}\right)\right|^2}
\label{eq:A5}
\end{equation}
\begin{equation}
E \bar{E}\left|\frac{1}{2} X+k \eta_2 Y\right|^2=\frac{\sqrt{2}}{k \eta_2}\left(1+\left(k \eta_2-\frac{1}{k \eta_2}\right)^2\right) \frac{1}{\left|\Gamma\left(\frac{3}{4}+\frac{i k^2}{2 \sigma^2}\right)\right|^2}
\label{eq:A6}
\end{equation}
\begin{equation}
\begin{aligned}
2 \sigma \eta_2 \operatorname{Re} e^{i \pi / 4} F \bar{E} X\left(\frac{1}{2} \bar{X}+k \eta_2 \bar{Y}\right) \\
&=\frac{2 \sigma}{\pi k}\left(\frac{1}{k^2 \eta_2^2} \exp \left(-\pi k^2 / 2 \sigma^2\right)-k \eta_2 \exp \left(\pi k^2 / 2 \sigma^2\right)\right)\\ \end{aligned}
\label{eq:A7}
\end{equation}

We now estimate the contribution of the $\Gamma$-functions. As indicated in the text, the two interesting ranges of values are $k / \sigma\ll1$ and $k / \sigma\gg 1$.
For $k / \sigma\ll1$, we use \cite{16}
\begin{equation}
|\Gamma(x+i y)|^2=\Gamma^2(x) \prod_{n=0}^{\infty}\left[\frac{1}{1+y^2 /(x+n)^2}\right]
\end{equation}

From which we learn that
\begin{equation}
\frac{1}{\left|\Gamma\left(\frac{1}{4}+\frac{i k^2}{2 \sigma^2}\right)\right|^2}=\frac{1}{\left(\Gamma\left(\frac{1}{4}\right)\right)^2}\left[1+O\left(\frac{k^4}{\sigma^4}\right)\right]
\end{equation}

Since $k / \sigma\ll1$ also means $k \eta_2<1$, we find the dominant contribution of 
\eqref{eq:A5} to be
\begin{equation}
F \bar{F} \text { term }: \frac{2 \sqrt{2}}{\left(\Gamma\left(\frac{1}{4}\right)\right)^2} \sigma^2 \eta_2^2 \frac{1}{k^3 \eta_2^3}\left(1+O\left(k^2 \eta_2^2\right)+O\left(\frac{k^4}{\sigma^4}\right)\right)
\end{equation}

Similarly, for 
\eqref{eq:A6} we find
\begin{equation}
E \bar{E} \text { term }: \frac{\sqrt{2}}{\left(\Gamma\left(\frac{3}{4}\right)\right)^2} \frac{1}{k^3 \eta_2^3}\left(1+O\left(k^2 \eta_2^2\right)+O\left(\frac{k^4}{\sigma^4}\right)\right)
\end{equation}

Finally, for the cross term 
\eqref{eq:A7} we get
\begin{equation}
F \bar{E} \text { term }: \frac{2}{\pi} \sigma \eta_2 \frac{1}{k^3 \eta_2^3}\left(1+O\left(k^3 \eta_2^3\right)+O\left(\frac{k^2}{\sigma^2}\right)\right)
\end{equation}

Comparing the three terms, we see that the $E \bar{E}$ term is the leading term of the expansion in $m / H$. Hence,
\begin{equation}
\left|\beta_k\right|^2 \simeq \frac{0.74}{\sqrt{2 m / H}\left(k_{p h y} / H\right)^3} \quad k_{p h y}^2\ll m H
\label{eq:A13}
\end{equation}
where $k_{p h y} \equiv k / \Omega_2$.
In the limit $k / \sigma\gg 1$, we use Stirling's approximation for the $\Gamma$-functions to find
\begin{equation}
|\Gamma(x+i y)|^2 \longrightarrow 2 \pi e^{-\pi |y|}|y|^{2 x-1} \quad(|y| \rightarrow \infty,|x| \ll |y|)
\label{eq:A14}
\end{equation}

Now the $\Gamma$-functions contribute important multiplicative factors $k / \sigma$, so that
\begin{equation}
F \bar{F} \text { term }: \frac{1}{\pi} \sigma \eta_2\left(1+\frac{1}{k^2 \eta_2^2}\right) \exp \left(\frac{\pi k^2}{2 \sigma^2}\right)
\end{equation}
\begin{eqnarray}
& E \bar{E} \text { term : } \frac{1}{\pi} \sigma \eta_2\left(1-\frac{1}{k^2 \eta_2^2}+\frac{1}{k^4 \eta_2^4}\right) \exp \left(\frac{\pi k^2}{2 \sigma^2}\right) \\
& F \bar{E} \text { term : } \quad-\frac{2}{\pi} \sigma \eta_2 \exp \left(\frac{\pi k^2}{2 \sigma^2}\right)
\end{eqnarray}

Since this range of $k$ corresponds to $k_{p h y} \sim H$, we have retained all the powers of $k \eta_2=$ $k_{\text {phy }} / H$. Combining these, we find
\begin{equation}
\left|\beta_k\right|^2=\frac{1}{4}\left(\frac{H}{k_{p h y}}\right)^4 \quad k_{p h y} \leq  H
\label{eq:A18}
\end{equation}

We now estimate $w_k$ of 
\eqref{eq:modefrequency} for both these cases.
\begin{equation}
\begin{aligned}
w_k= & \frac{1}{\Omega}\left|\chi_k^{\prime}\right|^2-2 \frac{\Omega^{\prime}}{\Omega^2} \operatorname{Re} \chi_k^{\prime} \chi_k^* \\
& +\Omega\left(\frac{|\vec{k}|^2}{\Omega^2}+m^2+\left(\frac{\Omega^{\prime}}{\Omega^2}\right)^2\right)\left|\chi_k\right|^2
\label{eq:A19}
\end{aligned}
\end{equation}

Since we evaluate this immediately at the end of inflation, we shall take $\Omega^{\prime} / \Omega^2=H$.
Then for $k^2 / \sigma^2 \ll 1$,
\begin{eqnarray}
\left|\chi_k\right|^2 & =\frac{1}{\sigma} \exp \left(-\pi k^2 / 2 \sigma^2\right) \frac{\pi}{\sqrt{2}} \frac{1}{\left(\Gamma\left(\frac{3}{4}\right)\right)^2} \\
\left|\chi_k^{\prime}\right|^2 & =\sigma \exp \left(-\pi k^2 / 2 \sigma^2\right) \pi \sqrt{2} \frac{1}{\left(\Gamma\left(\frac{1}{4}\right)\right)^2} \\
\operatorname{Re} \chi_k^{\prime} \chi_k^* & =-\frac{1}{2} \exp \left(-\pi k^2 / \sigma^2\right)
\end{eqnarray}

On comparing we find $H^2\left|\chi_k\right|^2$ making the dominant contribution, so that
\begin{equation}
w_k \simeq 1.5 \frac{H}{\sqrt{2 m / H}}+O\left(\left(\frac{m}{H}\right)^0,\left(\frac{k}{\sigma}\right)^2\right) \quad k_{p h_y}^2 \ll m H
\label{eq:A23}
\end{equation}

For $k^2 / \sigma^2 \gg 1$, we use \eqref{eq:A14} to find that
\begin{equation}
\left|\chi_k\right|^2 \rightarrow \frac{1}{2 k}
\end{equation}
\begin{equation}
\left|\chi_k^{\prime}\right|^2 \rightarrow \frac{k}{2}
\end{equation}
and the cross term is exponentially suppressed. Hence
\begin{equation}
w_k=k_{p h y}+\frac{H^2}{2 k_{p h y}}+O\left(\left(\frac{\sigma}{k}\right)^2\right) \quad \quad k_{p h y} \lesssim H
\label{eq:A26}
\end{equation}

\section{}\label{sec:appB}

Here we calculate the spectrum for the case $m / H=\sqrt{2}$. This makes the order of the Hankel functions $\lambda=1 / 2$. So
\begin{equation}
\begin{aligned}
X & =-i \sqrt{\frac{2}{\pi k \eta_2}} e^{i k \eta_2} \\
Y & =-i \sqrt{\frac{2}{\pi k \eta_2}}\left\{i k \eta_2-\frac{1}{2 k^2 \eta_2^2}\right\} e^{i k \eta_2}
\end{aligned}
\end{equation}

We have $\sigma \eta_2=\sqrt{2 m / H}=2^{3 / 4}$. Since we want the case $k_{p h y}^2 / \sigma^2 \ll 1$, we shall evaluate $D_{-\frac{1}{2}}\left(e^{i \pi / 4} \sigma \eta_2\right)$. Ideally we would like to expand in $k_{p h y}^2 / \sigma^2$ but this is not readily available, and as we shall see, not needed. We find the required values from the tables of Kireyeva and Karpov \cite{20} and using standard recursion relations for the Weber functions.
\begin{equation}
\begin{aligned}
& E=0.3960-i 0.6244 \\
& F=-0.0147+i 0.5565
\end{aligned}
\end{equation}

The $F \bar{F}$ term turns out to be the leading term in $k / \sigma$.
\begin{equation}
\left|\beta_k\right|^2=0.155 \frac{\sigma}{k}+0\left(\frac{k^2}{\sigma^2}\right)
\end{equation}

Next we calculate $w_k$ using
\begin{equation}
\left|\chi_k\right|^2=0.546 \frac{1}{\sigma} ; \quad\left|\chi_k^{\prime}\right|^2=0.31 \sigma
\end{equation}
\begin{equation}
\mathrm{Re}\chi_{k}'\chi_{k}^{*}\ =\ \mathrm{Re}e^{i\pi/4}F\bar{E}\ =\ -0.4
\end{equation}
\begin{equation}
w_k=2.3 H
\end{equation}
\begin{equation}
\frac{d\rho}{d k_{p h y}}~=~3.86~H^{2}k_{p h y}
\end{equation}

\end{document}